\documentclass[12pt]{article}
\begin{document}
\title{Superselection from canonical constraints\footnote{\copyright ~IOP Publishing 2004}}
\author{Michael J. W. Hall\\
Theoretical Physics, IAS, \\ Australian National
University,\\
Canberra ACT 0200, Australia}
\date{}
\maketitle



\begin{abstract}
The evolution of both quantum and classical ensembles may be described via the probability density $P$ on configuration space, its canonical conjugate $S$, and an {\it ensemble} Hamiltonian 
$\tilde{H}[P,S]$.  For quantum ensembles this evolution is, of course, equivalent to the Schr\"{o}dinger equation for the wavefunction, which is linear.  However, quite simple constraints on the canonical fields $P$ and $S$ correspond to {\it nonlinear} constraints on the wavefunction.  Such constraints act to prevent certain superpositions of wavefunctions from being realised, leading to superselection-type rules.  Examples leading to superselection for energy, spin-direction and `classicality' are given.  The canonical formulation of the equations of motion, in terms of a probability density and its conjugate, provides a universal language for describing classical and quantum ensembles on both continuous and discrete configuration spaces, and is briefly reviewed in an appendix.
\end{abstract}


\newpage

\section{Introduction}

The Schr\"{o}dinger equation for a quantum system is linear, implying that the superposition of any two solutions is also a solution.  However, some combinations of states have never been observed, including coherent superpositions of integer and half-integer spins, electric charges, and Schr\"{o}dinger's cat (wanted dead and alive).
Possible explanations for why such superpositions are not observed fall into two logical categories:\\
(i) {\it measurement superselection rules:} such superpositions may be allowed, but physical limitations on measurement prevent their observation;\\
(ii) {\it state superselection rules:} such superpositions are not physically allowed.\\ 
Each of these categories may of course be further subdivided into `in principle' and `in practice' subcategories.

A well known example of a {\it measurement} superselection rule is charge superselection in quantum field theory.  In particular, integration of the operator relation $\nabla .{\bf \hat{{\bf E}}}=4\pi\hat{\rho}$, relating the electric field and charge density, implies that the total charge operator $\hat{Q}=\int d^3x\,\hat{\rho}$ can be expressed as an integral over a surface containing any given finite region, and hence that all quasilocal field observables (those vanishing outside such regions) must commute with $\hat{Q}$ \cite{charge}.  Thus, the assumption that all physical observables are quasilocal implies that no measurement can distinguish relative phases of superpositions of different charge eigenstates, eg, between $|q_1\rangle + |q_2\rangle$ and $|q_1\rangle - |q_2\rangle$.  Note that this is not a {\it state} superselection rule, as such superpositions are not in themselves forbidden.  

Measurement superselection rules also include environment-induced superselection, forbidding the observation of coherent superpositions of classical apparatus states (and cats), due to decoherence arising from interaction between the apparatus and its (in practice unobservable) environment \cite{zurek, zeh,dice}.  It has been suggested that environment-induced superselection rules may also provide an alternative basis for explaining charge superselection \cite{giulini}.

State superselection rules are stronger than (and imply) measurement superselection rules (one cannot observe what does not exist).  They have typically been obtained by appealing to group symmetry arguments.  For example, assuming invariance of the wavefunction (up to a phase) under rotations of $2\pi$, the nontrivial nature of projective representations of the rotation group implies that superpositions of states of integer and half-integer spin are not permitted 
\cite{www,roman,galindo}.  Note that requiring a similar invariance for the wavefunctional of a charged field, under global phase shifts of the field, implies that superpositions of states of different total charge are not permitted.  However, as noted by Weinberg (his italics): ``{\it it may or may not be possible to prepare physical systems in arbitrary superpositions of states, but one cannot settle the question by reference to symmetry principles, because whatever one thinks the symmetry group of nature may be, there is always another group whose consequences are identical except for the absence of superselection rules}'' \cite{weinbook} (see also Ref.~\cite{giulini}).  This emphasises the fact that, in the end, superselection rules are incorporated into quantum theory via assumptions rather than as strict consequences.

The aim of this note is to point out an alternative mechanism for state superselection rules: nonlinear constraints on the wavefunction.  
Consideration of such constraints is perfectly natural - and in principle unavoidable - when one places physical significance on the existence of an action principle for quantum ensembles.  In particular, any Hamiltonian formulation of such an action principle requires two canonically conjugate fields on configuration space (eg, $\psi$ and $\psi^*$ \cite{roman, weinberg}, ${\rm Re}\,\psi$ and ${\rm Im}\,\psi$ \cite{heslot}, or $|\psi|$ and $\arg \psi$ \cite{eup,hkr,mehra}).  Specifying any constraints on these fields is then merely part of the general theory of Hamiltonian systems (that underlies, for example, classical mechanics, quantization on curved spaces, and standard discussions of quantum gauge symmetries \cite{teitel}).  Note that it is not convincing to argue that any such constraints must in fact be {\it linear} in $\psi$, on the grounds that the set of physical wavefunctions should be closed under superposition, as this argument amounts to simply assuming that there are no state superselection rules.

There is another fundamental justification for considering constraints that are nonlinear with respect to the wavefunction.  In particular, as will be reviewed in section 2, the evolution of both classical and quantum ensembles may be described in terms of an action principle for two canonically conjugate fields $P$ and $S$, where $P$ denotes the probability density on the configuration space.  In this common formalism, the primary differences between the classical and quantum cases are due to different ensemble Hamiltonians.  In particular, there is {\it no} difference in the treatment of  quantum and classical constraints: these are expressed in precisely the same way, in terms of the canonical fields $P$ and $S$.  Hence, since for classical ensembles one cannot even define, much less restrict to, the notion of constraints linear in some `wavefunction', it is rather difficult to justify any such {\it a priori} restriction for quantum ensembles.

It turns out that quite simple constraints on $P$ and $S$ can lead to interesting state superselection rules for quantum ensembles.  Moreover, such superselection rules can be qualitatively rather different from those which have been previously considered.  For example, they can act to not only restrict quantum ensembles to members of a set of orthogonal subspaces in Hilbert space, but to further permit only certain types of superpositions within these subspaces.  Indeed, they can even act to restrict quantum ensembles to members of a set of {non-orthogonal} states.  It should be emphasised that in all cases the evolution of the system is still linear, being described by the usual Schr\"{o}dinger equation.  

In the following section the Hamiltonian formulation of the equations of motion for classical and quantum ensembles of particles is briefly described. Differences between canonical constraints on such ensembles, and Dirac-type constraints on the wavefunction, are discussed in section 3.  Examples of constraints leading to energy, spin-direction and Gaussian superselection rules are given in section 4, and conclusions presented in section 5.

The Hamiltonian formulation of the description of ensembles on configuration space is of some interest in its own right, being quite universal in nature.  It is applicable to both continuous and discrete configuration spaces; to both classical and quantum ensembles; and even to non-orthogonal representations of quantum ensembles (eg, 
coherent-state representations).  Some general features of this formulation are therefore collected in an appendix.

\section{Canonical equations of motion}

The existence of an action principle for the evolution of a statistical ensemble, in a continuous configuration space, implies equations of motion of the Hamiltonian form (see appendix)
\begin{equation} \label{canon}
\frac{\partial P}{\partial t} = \frac{\delta \tilde{H}}{\delta S},~~~~~~~~\frac{\partial S}{\partial t} = -\frac{\delta \tilde{H}}{\delta P},  
\end{equation}
Here $P(x,t)$ denotes the probability density on the configuration space, $S(x,t)$ is the quantity canonically conjugate to $P$, and $\tilde{H}[P,S]$ is the {\it ensemble Hamiltonian}. The notation $\delta/\delta P$, $\delta/\delta S$ denotes functional derivatives with respect to $P$ and $S$ respectively (see appendix).

For example, for a classical ensemble of particles of mass $m$, acted on by an external potential $V(x)$, define the ensemble Hamiltonian 
\begin{equation} \label{hamc} 
\tilde{H}_c[P,S] :=  \int dx\,P\,\left[ \frac{|\nabla S|^2}{2m} + V(x) \right] .
\end{equation}
The equations of motion then follow via Eq.~(\ref{canon}) and 
Eq.~(\ref{funcderiv}) of the appendix as
\begin{equation} \label{conthj}
\frac{\partial P}{\partial t} + \nabla .(P\nabla S/m) = 0,~~~~~~
\frac{\partial S}{\partial t} + \frac{|\nabla S|^2}{2m} + V = 0 .
\end{equation}
The first of these is a continuity equation, ensuring conservation of probability, while the second is the classical Hamilton-Jacobi equation \cite{goldstein}.  Note that if $\nabla S$ is interpreted as the momentum of a particle at position $x$, then the classical ensemble Hamiltonian $\tilde{H}_c$ corresponds to the average ensemble energy.  However, such an interpretation will not be relied upon here.

As a second example, for a {\it quantum} ensemble of such particles define the ensemble Hamiltonian
\begin{equation} \label{hamq} 
\tilde{H}_q[P,S] :=  \tilde{H}_c[P,S] + \frac{\hbar^2}{4} \int dx\,P \frac{|\nabla \log P|^2}{2m} . 
\end{equation}
The corresponding equations of motion for $P$ and $S$ follow as
\begin{equation} \label{qconthj}
\frac{\partial P}{\partial t} + \nabla .(P\nabla S/m) = 0,~~~~~~
\frac{\partial S}{\partial t} + \frac{|\nabla S|^2}{2m} + V +Q= 0 ,
\end{equation}
where $Q:=-(\hbar^2/2m)P^{-1/2}\nabla^2P^{1/2}$ is the so-called `quantum potential' \cite{holland}.  It is well known that if one defines $\psi:=P^{1/2}e^{iS/\hbar}$, these equations of motion are equivalent to the Schr\"{o}dinger equation
\begin{equation} \label{se}
i\hbar \frac{\partial\psi}{\partial t}= \frac{-\hbar^2}{2m}\nabla^2 \psi + V\psi.
\end{equation}
It may be checked that the quantum ensemble Hamiltonian $\tilde{H}_q$ is equal to the average quantum ensemble energy $\int dx\, \psi^*(-\hbar^2\nabla^2/2m +V)\psi$ (see also appendix).  Note also that $\tilde{H}_q$ reduces to $\tilde{H}_c$ in the classical limit $\hbar\rightarrow 0$.  

These examples indicate that both classical and quantum ensembles may be described via two canonically conjugate fields $P$ and $S$ on configuration space, with differences in evolution corresponding to differences in the respective ensemble Hamiltonians $\tilde{H}_c$ and $\tilde{H}_q$.  Note from Eq.~(\ref{hamq}) that these ensemble Hamiltonians differ by a nonclassical kinetic energy term, inversely proportional to mass $m$. 
Some general properties of the canonical formulation for ensembles on configuration space (whether classical or quantum, continuous or discrete) are discussed in the appendix.
 
\section{Dirac constraints vs canonical constraints}

The  canonical formulation of the equations of motion provides a common basis for discussing both classical and quantum ensembles, and in particular for treating classical and quantum constraints on an equal footing.  More significantly, for the purposes of this note, the canonical formulation in fact leads to a more general treatment of quantum constraints than does the standard approach introduced by Dirac \cite{dirac}.

In particular, the standard approach relies on replacing classical constraints of the form $C(x,p)=0$ by  operator constraints on the wavefunction, of the form \cite{weinbook, teitel,dirac}
\begin{equation} \label{dirac}
C(x, \frac{\hbar}{i}\nabla) \psi =0  .
\end{equation}
Consistency of these constraints with the equations of motion and with each other then leads to a number of secondary constraints of a similar form.  What is important to note here is that in this approach all constraints are {\it linear} with respect to the wavefunction.  Hence any superposition of two wavefunctions, each satisfying the constraints, will also satisfy the constraints, i.e., {\it the standard approach cannot lead to superselection rules}.

In contrast, in the canonical formulation one is dealing, in both the quantum and the classical cases, with a Hamiltonian system for two conjugate fields, $P$ and $S$.  Hence, any primary constraints are expressed in precisely the {\it same} manner, 
\begin{equation} \label{cancon}
K[P,S]=0, 
\end{equation} 
in both cases \cite{hkr}.  Any fundamental classical/quantum differences in this regard arise because the corresponding ensembles have different ensemble Hamiltonians, so that any {\it secondary} constraints, following from consistency with the equations of motion, will in general be different \cite{teitel}.  It may be assumed that the functional forms of the classical and quantum primary constraints are identical, i.e., $K_q\equiv K_c$ (this is the simplest possible correspondence).  Note, however, that physical consistency only requires identity to lowest order in $\hbar$.

Now, every Dirac-type constraint on the wavefunction, of the form of Eq.~(\ref{dirac}), clearly corresponds to {\it two} canonical constraints on the fields $P$ and $S$, of the form of 
Eq.~(\ref{cancon}) (obtained by substituting $\psi=P^{1/2}e^{iS/\hbar}$ in Eq.~(\ref{dirac}) and taking real and imaginary parts).  For example, the Dirac constraint $(x\times\nabla)\psi=0$ (rotational symmetry of $\psi$) maps to the corresponding canonical constraints $x\times\nabla P=0$, $x\times\nabla S=0$ (rotational symmetry of $P$ and $S$ respectively).  More general symmetries of $\psi$ also map to two corresponding symmetries for $P$ and $S$ (eg, invariance of $\psi$ under rotation by $2\pi$, up to a phase factor, maps to the invariance of $P$, and to the invariance of $S$ up to an additive constant).

However, the set of possible canonical constraints is much larger than those corresponding to Dirac-type constraints, and in particular includes constraints on $P$ and $S$ that correspond to {\it nonlinear} constraints on $\psi$ (examples are given in the next section).  Such nonlinear constraints are not invariant under superposition, and  hence lead directly to state superselection rules for quantum ensembles.

As remarked in the Introduction, the possibility of {\it classical} constraints which are not linear for the particular combination of canonical fields $\psi=P^{1/2}e^{iS/\hbar}$ is not at all surprising (and rather desirable!).  Yet the canonical formalism applies equally well to both classical and quantum ensembles.  Imposing a restriction to linear constraints, for quantum ensembles only, is thus rather difficult to justify on physical grounds, and the full set of possible canonical constraints therefore deserves serious consideration.

\section{Superselection via canonical constraints}

It has been shown that the set of canonical constraints for configuration-space ensembles includes those corresponding to {\it nonlinear} constraints on the wavefunction.  Thus, although solutions of the Schr\"{o}dinger equation Eq.~(\ref{se}) are closed under arbitrary superpositions, the presence of such constraints in general acts to rule out a number of such superpositions, leading directly to superselection-type rules.  

Several examples are given here, related to superselection of energy, spin direction, and `classicality'.  Note that these examples are intended only to be indicative of the manner in which nonlinear constraints can lead to state superselection rules.  Further investigation is needed to determine whether well-motivated constraints exist that lead to superselection rules of physical interest.

\subsection{Energy superselection}

Consider first the rather simple canonical constraint
\begin{equation} \label{simple}
 J := P\,\nabla S = 0 .
\end{equation}
This constraint is local, invariant under $S\rightarrow S+{\rm constant}$, and may be physically interpreted as the requirement that the ensemble momentum density vanishes everywhere (see also appendix).  Note that for quantum ensembles it can be re-expressed as ${\rm Im}\, \psi^*\nabla\psi=0$, which clearly cannot be put in the linear Dirac form of Eq.~(\ref{dirac}).

To investigate this constraint for a {\it classical} ensemble of particles, note that consistency with the equations of motion requires $\partial J/\partial t =0$ \cite{weinbook, teitel, dirac}.  
Eqs.~(\ref{conthj}) and (\ref{simple}) then yield the secondary constraint
\begin{eqnarray}
0 = \partial (P\nabla S)/\partial t &=& (\partial P/\partial t)\nabla S + P (\partial (\nabla S)/\partial t)\nonumber\\
&=& -[\nabla .(P\nabla S)]\nabla S - P\,\nabla[|\nabla S|^2/(2m) + V]\nonumber \\ \label{secc}&=& -P\nabla V .
\end{eqnarray}
Hence the classical force, $-\nabla V$, vanishes over the suport of the ensemble.  Note in particular that if the potential energy has a single minimum, then the constraint requires the ensemble to be concentrated solely at this minimum, i.e.,  the ensemble must occupy the classical groundstate.  

In contrast, for a {\it quantum} ensemble, Eq.~(\ref{simple}) requires that $\nabla S$ vanishes on the support of the wavefunction, and hence that $S$ has no spatial dependence for $P\neq0$.  Secondary constraints arising from consistency with the equations of motion can be determined similarly to the classical case above (one finds that the `quantum' force, $-\nabla(V+Q)$, must vanish over the ensemble, where $Q$ is the 
quantum potential in Eq.~(\ref{qconthj})). However, it is simpler to directly substitute the ansatz $S(x,t)=-f(t)$ into the Schr\"{o}dinger equation Eq.~(\ref{se}) and use the continuity equation (\ref{qconthj}), to obtain the secondary constraints
\[ \dot{f}P^{1/2} = \left[\frac{-\hbar^2}{2m}\nabla^2 + V\right] 
P^{1/2},~~~~~~\partial P/\partial t=0 \]
respectively.  Differentiating the first of these with respect to time  and applying the second implies that $\dot{f}=E={\rm constant}$, and hence these constraints are equivalent to the time-independent Schr\"{o}dinger equation
\begin{equation} \label{secq}
\frac{-\hbar^2}{2m}\nabla^2 \psi + V\psi= E\psi .
\end{equation}
Thus the quantum ensemble is required to be in an energy eigenstate (but not necessarily in the quantum groundstate).

It is seen that in both the classical and quantum cases, the primary constraint in Eq.~(\ref{simple}) leads to the requirement that the ensemble is stationary.  In the quantum case this immediately yields a state superselection rule: {\it superpositions of states of different energy are forbidden}.  Thus this constraint provides a very simple example of how canonical constraints can lead to superselection-type rules for quantum ensembles.  

Note that for degenerate Hamiltonians the above example is even more stringent than usual superselection rules. In particular, if $\psi_1$ and $\psi_2$ are two real-valued energy eigenstates corresponding to the {\it same} energy eigenvalue $E$, then it follows from Eq.~(\ref{simple}) that only superpositions of the form
\[ e^{i\phi}\left( a\psi_1 + b\psi_2\right) ,\]
are permitted, where $a$ and $b$ are real.  Thus the physical states are not only restricted to members of a set of orthogonal subspaces of the Hilbert space;  they are further restricted, up to an arbitrary phase factor, to {\it real} superpositions of (real-valued) members of these subspaces.  The constraint thus effectively lowers the degeneracy of the Hamiltonian (it would be interesting to investigate the thermodynamic implications).

\subsection{Spin-direction superselection}

As an example involving a {\it discrete} configuration space, an ensemble of two-level systems will be considered. The configuration space is therefore labelled by two values, $j=1,2$, and the canonical equations of motion have the form (see appendix)
\begin{equation} \label{canond}
\dot{P_j} = \frac{\partial\tilde{H}}{\partial S_j},~~~~~~\dot{S_j} = -
\frac{\partial\tilde{H}}{\partial P_j} 
\end{equation}
for some ensemble Hamiltonian $\tilde{H}(P_1,P_2,S_1,S_2)$.  

For the particular example of a quantum ensemble of spin-half particles, in an external magnetic field ${\bf B}$, the ensemble Hamiltonian has the form (see appendix)
\begin{eqnarray*} 
\tilde{H}(P_1,P_2,S_1,S_2) &:=& \mu\langle \sigma.{\bf B}\rangle\\ &=& 
\mu(P_1-P_2)B_z\\&~&~~ + 2\mu\sqrt{P_1P_2}\left[B_x\cos\frac{S_1-S_2}{\hbar}-B_y \sin\frac{S_1-S_2}{\hbar}\right] ,
\end{eqnarray*}
where $\mu$ is the magnetic moment; $\sigma$ denotes the Pauli sigma matrices; and $B_x$, $B_y$ and $B_z$ denote the components of ${\bf B}$ relative to some set of orthogonal directions $x,y,z$.  The canonical coordinates are related to the usual Bloch sphere representation by $P_1=\cos^2\theta/2$, $S_1-S_2=\phi$, and it is well known that evolution corresponds to rotation of the Bloch sphere about an axis parallel to the magnetic field ${\bf B}$.  Such rotations thus correspond to a particular group of canonical transformations of the canonical coordinates $P_1$, $P_2$, $S_1$ and $S_2$.

Consider now the constraint that 
\begin{equation} \label{spin}
S_1 - S_2 = 0 
\end{equation}
for some choice of $x,y,z$.  This constraint is analogous to Eq.~(\ref{simple}), and is similarly invariant under the transformation $S_j\rightarrow S_j+C$ (and hence consistent with the conservation of probability).  It restricts the ensemble to lie on a great circle on the Bloch sphere (passing through the $z$-axis), and thus geometrically corresponds to a rather natural {\it geodesic} constraint in the Bloch representation.  Note, however, that it cannot be expressed as a Dirac-type constraint on the wavefunction.

The secondary constraints arising from the requirement of compatibility of Eq.~(\ref{spin}) with the equations of motion follow directly from geometrical considerations:  rotation of the Bloch sphere about ${\bf B}$ will leave the ensemble on some great circle if and only if (i) the axis of rotation is orthogonal to the great circle, or (ii) the ensemble lies on the intersection of the great circle with the axis of rotation ${\bf B}$.  

The constraint thus leads to a state superselection rule for spin direction: the ensemble either corresponds to one of the two eigenstates of $\sigma.{\bf B}$, or to an equally-weighted-modulus superposition of these eigenstates.  Note that this superselection rule is consistent with measurement of spin via a Stern-Gerlach apparatus, which leaves the spin-direction either parallel or antiparallel to the magnetic field of the apparatus.

\subsection{Gaussian superselection}

The ensemble Hamiltonians $\tilde{H}_c$ and $\tilde{H}_q$, for classical and quantum ensembles of particles respectively, differ by a functional of the position probability density $P$, as indicated in Eq.~(\ref{hamq}).  This functional vanishes in the formal limit $\hbar\rightarrow 0$, as one might expect.  However, in the real world one cannot take such a limit - $\hbar$ is a fixed and fundamental constant - and hence any physical description of a classical limit for quantum ensembles must take a different approach \cite{jones}.

Here one very simple criterion for quantum ensembles to behave classically is considered:  {\it the difference between the quantum and classical ensemble Hamiltonians should be as small as possible}.  It follows from Eq.~(\ref{hamq}) that this criterion is equivalent to the constraint 
\begin{equation} \label{classcon}
F := \int dx \, P\,|\nabla \log P|^2 = {\rm minimum}.
\end{equation}
Note that this cannot be expressed as a Dirac-type constraint on the wavefuction. 

The quantity $F$ in Eq.~(\ref{classcon}) is just the (classical) Fisher information of $P$ \cite{cover}.  It is strictly positive, and satisfies the isoperimetric inequality 
\[ F\geq (2\pi en) e^{-2H_X/n}, \]
where $H_X$ denotes the entropy of the ensemble and $n$ is the dimension of the configuration space, with equality holding if and only $P$ is {\it Gaussian} \cite{cover}.  It follows immediately, under the additional (rather weak) assumption that the entropy of $P$ is finite, that the classicality constraint in Eq.~(\ref{classcon}) is automatically satisfied if $P$ can be chosen to be Gaussian, i.e., of the form
\begin{equation} \label{gauss}
P(x,t) = (2\pi e)^{-n/2}(\det K_t)^{1/2}e^{-(1/2)(x-a_t)^TK_t(x-a_t)}
\end{equation}
for some (possibly time-dependent) non-singular matrix $K_t$ and mean $a_t$.  Note that the finite entropy assumption holds, for example, whenever the configuration space variances ${\rm Var}(x_1), {\rm Var}(x_2), ...$ are finite.

Substitution of Eq.~(\ref{gauss}) into Eqs.~(\ref{qconthj}) leads in a straightforward manner to the secondary constraints  that each of $S$ and $V$ are at most quadratic in $x$.  Hence, for {\it quadratic} Hamiltonians (eg, the harmonic oscillator, a particle in a constant magnetic or gravitational field, and free particles), {\it the classicality constraint} (\ref{classcon}) {\it requires the wavefunction to be a complex Gaussian in} $x$.  

The classicality constraint thus leads to a rather strong state superselection rule for systems with quadratic Hamiltonians (and finite position entropy): the wavefunction is restricted to a set of {\it non-orthogonal} states.  The connection between such `classical' wavefunctions and classical trajectories has been recently reviewed in \cite{andrews}.

\section{Conclusions}

State superselection rules and nonlinear constraints on the wavefunction are inextricably linked in a formal sense.  For example, any superselection rule of the form that physical wavefunctions are restricted to a set of orthogonal subspaces, corresponding to an orthogonal set of projections $\{\hat{E}_j\}$ (and hence to eigenstates of a Hermitian operator $\hat{A}:=\sum_j a_j\hat{E}_j$), is formally equivalent to the nonlinear constraint
\begin{equation} \label{formula} 
\sum_j \langle \psi|E_j|\psi\rangle^2 =1  
\end{equation}
on the wavefunction.  It follows that if state superselection rules have physical content, then so must nonlinear constraints, and vice versa.  

It has been demonstrated, in the context of configuration-space ensembles satisfying an action principle, that the consideration of nonlinear constraints is in principle unavoidable.  Moreover, in this context both quantum and classical constraints can be discussed on an equal footing, generalising the standard approach based on mapping classical phase space constraints to linear operator constraints.

The examples of constraints leading to superselection of energy, spin direction, and `classicality' in section 4 show that quite simple nonlinear constraints on the wavefunction can lead to interesting state superselection rules.  However, further investigation is needed to determine whether well-motivated constraints exist (i.e., not of the {\it ad hoc} form of Eq.~(\ref{formula}) above), that lead to  superselection rules for quantities such as total charge and spin discussed in the Introduction.  It would also be of interest to investigate what conditions need to be imposed, if any, to ensure the consistency of such superselection rules with physical measurements.

The canonical formulation of the equations of motion emphasises the fact that an action principle for quantum ensembles requires {\it two} fields.  These are usually taken as $\psi$ and $\psi^*$ \cite{roman}, but of course can also be taken as $P$ and $S$.  Interpretations of quantum mechanics which neglect this fact (i.e., which neglect to provide a physical interpretation for both $\psi$ and $\psi^*$, or both $P$ and $S$), correspond to according the quantum equations of motion primary significance over the action principle from which they follow.  This is opposite to the usual point of view taken in {\it classical} dynamics \cite{goldstein}.

It is perhaps worth pointing out that, since the wavefunction can be recovered from tomographic measurements (up to a phase factor), the fields $P$ and $S$ can similarly be recovered (the latter up to an additive constant).  Indeed, weak measurements of momentum on an ensemble of particles, postselected by strong measurements of position, allow one to recover both $P(x)$ and $\nabla S$ for both the classical and the quantum case \cite{weak} (where the latter can be integrated to give $S$ up to a constant).

Finally, note that the canonical formulation of the equations of motion only applies to pure states in the quantum case.  It would be of interest, for thermodynamic and other purposes, to consider properties of statistical mixtures of configuration-space ensembles (corresponding to density operators in the quantum case).  \\

{\bf ACKNOWLEDGMENTS}\\
I am grateful to Marcel Reginatto and Howard Wiseman for helpful remarks.


\appendix
\section{Appendix: Configuration-space ensembles}

To introduce configuration-space ensembles at a fundamental level, suppose that (i) the configuration of a physical system can only be specified imprecisely, requiring it to be described by an ensemble on the configuration space, and (ii) there is a action principle for the motion of the ensemble.

The first assumption implies that the configuration of the physical system is formally described by a probability density $P$ on configuration space.  The second assumption implies the existence of some conjugate function $S$ on configuration space, and a `ensemble' Hamiltonian $\tilde{H}[P,S,t]$, such that the equations of motion are generated by Hamilton's action principle $\delta \alpha=0$, with \cite{goldstein}
\begin{equation} \label{action} 
\alpha:= \int dt \,\left[ \tilde{H} + \int P\frac{\partial S}{\partial t}\right] = \int dt\, \left[ \tilde{H}+\langle \partial S/\partial t\rangle \right] . 
\end{equation}
The inner integral in the first equality above is over the configuration space, and is replaced by summation over any discrete parts of configuration space.  The physical interpretation of $S$ will be discussed further below.  For continuous configuration spaces it is typically an energy-momentum potential.

It is perhaps simplest to first consider {\it discrete} configuration spaces (corresponding, eg, to the description of classical dice throws and quantum spins).  The probability of configuration $j$ is then $P_j$, with conjugate quantity $S_j$, and the equations of motion follow from Eq.~(\ref{action}) as
\begin{equation} \label{hamdis} 
\dot{P}_j = \partial \tilde{H}/\partial S_j,~~~~~~\dot{S}_j = -\partial \tilde{H}/\partial P_j . 
\end{equation}
Only Hamiltonians which conserve probability are of interest, and hence, to first order in $\epsilon$, it is required that
\[ 0=\epsilon\sum_j\dot{P}_j = \epsilon\sum_j\frac{\partial \tilde{H}}{\partial S_j} = \tilde{H}(P, S+\epsilon)- \tilde{H}(P,S) . \]
Thus, $\tilde{H}$ is invariant under the addition of a constant to the components of $S$.  

It follows for discrete configuration spaces that the ensemble Hamiltonian must be of the form
\begin{equation} \label{form}
\tilde{H} = F(P, M) ,
\end{equation}
where $M$ denotes the antisymmetric matrix with components $M_{jk}=S_j-S_k$.  The equation of motion for $P_j$ therefore reduces to the familiar rate equation form \cite{wiseman}
\begin{equation}  \label{rate}
\dot{P}_j = \sum_k \left( \frac{\partial F}{\partial M_{jk}} - 
\frac{\partial F}{\partial M_{kj}} \right)  = \sum_k \left( T_{jk}P_k - T_{kj} P_j \right) , 
\end{equation}
with transition rates
\begin{equation} \label{tran} 
T_{jk} := P_k^{-1} \frac{\partial F}{\partial M_{jk}} 
\end{equation}
(with $T_{jk}:=0$ for $P_k=0$).  Noting that  probability must remain positive under evolution implies further that
\begin{equation} \label{pos}
\left. \frac{\partial \tilde{H}}{\partial S_j}\right|_{P_j=0} = \left. \sum_{k(\neq j)} P_k\,T_{jk}\right|_{P_j=0} \geq 0 . 
\end{equation}
For constant transition rates this reduces to $T_{jk}\geq0$ for $j\neq k$.

For a {\it continuous} configuration space, indexed by some (possibly multidimensional) parameter $x$, the ensemble Hamiltonian becomes a {\it functional}, $\tilde{H}[P,S]$, and the equations of motion follow from Eq.~(\ref{action}) as
\[ \frac{\partial P}{\partial t} = \frac{\delta \tilde{H}}{\delta S},~~~~~~~~\frac{\partial S}{\partial t} = -\frac{\delta \tilde{H}}{\delta P} . \]
Here $\delta F/\delta f$ denotes the functional derivative of functional $F[f]$ with respect to function $f$, where for the special case that $F[f] = \int dx\, R(f,\nabla f,x)$ one has \cite{goldstein}
\begin{equation} \label{funcderiv} 
\delta F/\delta f = \partial R/\partial f - \nabla.[\partial R/\partial (\nabla f)] .
\end{equation}

The conservation and positivity of probability places restrictions on the ensemble Hamiltonian analogous to the discrete case above.  In particular, conservation of probability implies that $\tilde{H}$ is invariant under $S(x)\rightarrow S(x)+{\rm constant}$, and so can only depend on $\nabla S$ and higher derivatives.  The equation of motion for $P$ can therefore always be expressed in the form of a continuity equation, 
\[ \partial P/\partial t + \nabla .J=0, \] 
for some current $J$ (one formally has $J=\delta \tilde{H}/\delta(\nabla S)$).  

Examples of ensemble Hamiltonians for classical and quantum particles have been given in Eqs.~(\ref{hamc}) and (\ref{hamq}) respectively (with current $J=P\nabla S/m$ in both cases).  The form of the classical Hamiltonian $\tilde{H}_c$ is perhaps the simplest possible scalar functional of $\nabla S$, and is well known (dating back to the 19th century) from the theory of ideal fluids \cite{fluid}.  The form of the quantum Hamiltonian $\tilde{H}_q$ may be derived from the corresponding classical form via an `exact uncertainty' principle, both for particles \cite{eup} and for bosonic fields \cite{hkr}.  

It is straightforward to show that {\it all} quantum systems may be formulated in terms of configuration-space ensembles, where the `configuration space' may be chosen as corresponding to the outcome space of any complete measurement on the system.  In particular, let $\hat{H}$ be the Hamiltonian operator for a quantum system described by state $|\psi\rangle$, and let $\{|a\rangle\}$ denote any  basis set satisfying the completeness property
\[ \int da\, |a\rangle\langle a| = \hat{1}  \]
(with integration replaced by summation over any discrete ranges of $a$).  Such a basis set corresponds to the outcomes associated with the measurement of some observable $A$ (eg, position, momentum, optical phase, energy, spin, etc.), where $|\langle a|\psi\rangle|^2$ is the probability density associated with measurement outcome $a$.  Note it is {\it not} required that the basis set be orthonormal.

Now define $P(a)$ and $S(a)$ by the polar decomposition $\langle a|\psi\rangle = P^{1/2} e^{iS/\hbar}$, and the associated ensemble Hamiltonian $\tilde{H}_A$ by
\begin{equation} \label{hama} 
\tilde{H}_A[P,S] = \langle \psi|\hat{H}|\psi\rangle = \int da\, da'\, [P(a)P(a')]^{1/2}\langle a|\hat{H}|a'\rangle e^{i[S(a')-S(a)]/\hbar} .
\end{equation}
It follows that the action in Eq.~(\ref{action}) can be rewritten as
\[ \alpha = \int dt\, \left[ \langle \psi|\hat{H}|\psi\rangle + 
\frac{1}{2}i\hbar\int da\, \left( \frac{ \partial\langle a|\psi\rangle^*}{ \partial t}\, \langle a|\psi\rangle - \langle a|\psi\rangle^* 
\frac{ \partial\langle a|\psi\rangle}{ \partial t}\right) \right] ,\]
and hence (by variation with respect to $\langle a|\psi\rangle$ and its conjugate) that the equations of motion for $P$ and $S$ are identical to the Schr\"{o}dinger equation (and its conjugate) in the 
$a$-representation.  

The natural form of the transition rates for a {\it discrete} quantum observable $A$ follows from Eqs.~(\ref{tran}) and (\ref{hama}) as 
\begin{equation} 
T_{jk} = \hbar^{-1}(P_j/P_k)^{1/2}\,{\rm Im}\,\left\{ \langle a_j|
\hat{H}|a_k\rangle e^{-i(S_j-S_k)/\hbar}\right\} . 
\end{equation}
The positivity condition (\ref{pos}) automatically follows (with equality) from the Hermiticity of $\hat{H}$.  An alternative `geometric'  approach to the configuration-space ensemble formulation, for the case of discrete quantum ensembles, has been given recently by Mehrafarin \cite{mehra}.  

It is of interest to give an explicit example of the quantum equations of motion for the case of a {\it non-orthogonal} basis set (see also \cite{wiseman}).  In particular, consider a single-mode bosonic field with Hamiltonian operator $\hat{H}=\hbar\omega \hat{N}$.  The phase observable $\Phi$ then corresponds to the basis set $\{|\phi\rangle\}$ with number-state expansion \cite{helstrom, shapiro}
\[ |\phi\rangle := (2\pi)^{-1/2}\sum_{n=0}^{\infty} e^{in\phi}|n\rangle . \]
It follows that $\langle\phi|\hat{N}|\psi\rangle = i(d/d\phi) \langle \phi|\psi\rangle$, and hence that the corresponding ensemble Hamiltonian is given by 
\[ \tilde{H}_\Phi[P,S] = \hbar\omega\langle\psi|\hat{N}|\psi\rangle = -\omega \int_0^{2\pi} d\phi\, P\frac{\partial S}{\partial\phi} . \]
The associated equations of motion follow as
\[ \partial P/\partial t = \omega(\partial P/\partial\phi),  ~~~~~~\partial S/\partial t = \omega(\partial S/\partial\phi) .\]
These may be solved directly, or by noting that $\tilde{H}_\Phi$ is proportional to the ensemble momentum associated with phase translations (cf. Eq.~(\ref{linmom}) below), to give $P(\phi,t)=P_0(\phi +\omega t)$ and $S(\phi, t)=S_0(\phi+\omega t)$.

Finally, to establish the physical interpretation of the conjugate quantity $S$, recall from the above discussion that the conservation of probability implies invariance of $\tilde{H}$ under the transformation $S\rightarrow S+C$. The equations of motion are also invariant under this transformation.  Hence only derivatives and relative changes in $S$ are expected to have direct physical significance.  

For example, the form of the ensemble action in Eq.~(\ref{action}) suggests that $\partial S/\partial t$ is a local energy associated with the ensemble.  This can be made more precise, both for continuous and discrete configuration spaces, in the case that the ensemble Hamiltonian is {\it homogeneous} with respect to $P$, i.e.,
\begin{equation} \label{homog} 
\tilde{H}[\lambda P, S] = \lambda^r \tilde{H}[P,S] 
\end{equation}
for all $\lambda>0$ and some constant $r$.  
In particular, differentiating the homogeneity condition with respect to $\lambda$ and putting $\lambda=1$ yields the identity
\begin{equation} 
\tilde{H} = r^{-1} \int dx \,P\frac{\delta\tilde{H}}{\delta P} = -r^{-1} \int dx\, P\frac{\partial S}{\partial t} = -r^{-1} \langle \partial S/\partial t \rangle  
\end{equation}
(integration is replaced by summation for discrete configuration spaces).  Thus, identifying $\tilde{H}$ with the ensemble energy (which is certainly justified for time-independent Hamiltonians), it follows that $-r^{-1}P(\partial S/\partial t)$ {\it may be interpreted as a local energy density}.  This is further supported by the invariance of the equations of motion under $\tilde{H}\rightarrow\tilde{H}+\epsilon(\int dx\,P)^r$, $S \rightarrow S-\epsilon rt$.  Note that the homogeneity property (\ref{homog}) in fact holds for all cases considered in this paper, with $r=1$.  

Further, for continuous configuration spaces, consider an infinitesimal translation in configuration space.  The change in the ensemble Hamiltonian is then, to first-order,
\begin{eqnarray*} 
\tilde{H}[P(x+\delta x),S(x+\delta x)]-\tilde{H}[P(x),S(x)] &=& \delta x.\int dx\, \left( \frac{\delta\tilde{H}}{\delta P}\nabla P +\frac{\delta\tilde{H}}{\delta S}\nabla S\right)\\ &=& \delta x.\frac{\partial}{\partial t} \int dx\, P\nabla S ,
\end{eqnarray*}
where the last equality follows using the equations of motion and integration by parts.  It follows that the linear momentum of the ensemble may be identified as \cite{goldstein}
\begin{equation} \label{linmom} 
\Pi = \int dx\, P\nabla S = \langle \nabla S \rangle ,
\end{equation}
and hence that $P\nabla S$ {\it may be interpreted as a local momentum density associated with the ensemble}.


\begin{thebibliography}{99}

\bibitem{charge} Strocchi F and Wightmann A S 1974 {\it J. Math. Phys.} {\bf 15} 2198
\bibitem{zurek}  Zurek W H 1982  {\it Phys. Rev. D} {\bf 28} 1862
\bibitem{zeh} Zeh H D 1993 {\it Phys. Lett. A} {\bf 172} 189
\bibitem{dice} Elze H-T (ed) 2004 {\it Decoherence and Entropy in Complex Systems} (Berlin: Springer)
\bibitem{giulini} D Giulini 2000 {\it Decoherence:Theoretical, Experimental, and Conceptual Problems} ed P Blanchard, D Giulini, E Joos, C Kiefer and I-O Stamatescu (Berlin: Springer) pp. 87-100; {\it Preprint} quant-ph/9906108
\bibitem{www}  Wick G S, Wightmann A S and Wigner E P 1952 {\it Phys. Rev.} {\bf 88} 101
\bibitem{roman} Roman P 1965 {\it Advanced Quantum Theory} (Reading: Addison-Wesley) pp 31-33, 65, 539
\bibitem{galindo} Galindo A and Pascual P 1990 {\it Quantum Mechanics I} (Berlin: Springer) pp 85-87, 209, 292-293
\bibitem{weinbook} Weinberg S 1995 {\it The Quantum Theory of Fields Volume 1} (Cambridge: Cambridge University Press) pp 21-22, 90-91, 325-331
\bibitem{weinberg} Weinberg S 1989 {\it Ann. Phys. (NY)} {\bf 194} 336
\bibitem{heslot} Heslot A 1985 {\it Phys. Rev. D} {\bf 31} 1341
\bibitem{eup} Hall M J W and Reginatto M 2002 {\it J. Phys. A} {\bf 35} 3289
\bibitem{hkr} Hall M J W, Kumar K and Reginatto M 2003 {\it J. Phys. A} {\bf 36} 9779
\bibitem{mehra} Mehrafarin M 2004 {\it Preprint} quant-ph/0402153
\bibitem{teitel} Henneaux M and Teitelboim C 1992 {\it Quantization of Gauge Systems} (New Jersey: Princeton) ch 1
\bibitem{goldstein} Goldstein H 1950 {\it Classical Mechanics} (New York: Addison-Wesley) ch. 7, 9, 11
\bibitem{holland} Holland P R 1993 {\it The Quantum Theory of Motion} (Cambridge: Cambridge University Press) \bibitem{dirac} Dirac P A M 1966 {\it Lectures on Quantum Field Theory} (New York: Academic) ch 14, 15
\bibitem{jones} Bhattacharya T, Habib S and Jacobs K 2000 {\it Phys. Rev. Lett.} {\bf 85} 4852
\bibitem{cover} Dembo A, Cover T M and Thomas J A 1991 {\it IEEE Trans. Inf. Theory} {\bf 37} 1501
\bibitem{andrews} Andrews M 1999 {\it Am. J. Phys.} {\bf 67} 336
\bibitem{weak} Aharonov Y, Albert D Z and Vaidman L 1988 {\it Phys. Rev. Lett.} {\bf 60} 1351
\bibitem{wiseman} Gambetta J and Wiseman H M 2004 {\it Found. Phys.} {\bf 34} 419
\bibitem{fluid} Zakharov V E and Kuznetsov E A 1997 {\it Physics - Uspekhi} {\bf 40} 1087
\bibitem{helstrom} Helstrom C W 1974 {\it Int. J. Theoret. Phys.} {\bf 11} 357
\bibitem{shapiro} Shapiro J H and Shepard S R 1991 {\it Phys. Rev. A} {\bf 43} 3818

\end{thebibliography}
\end{document}